\documentclass[final,3p,times,sort&compress]{elsarticle}

\usepackage{amsmath}
\usepackage{amssymb}
\usepackage{graphicx}
\usepackage{dcolumn}
\usepackage{bm}
\usepackage{mathptmx}
\usepackage{tikz}
\usepackage[subfigure]{graphfig}
\usepackage{float}
\usepackage{dsfont}
\usepackage{color}
\usepackage{caption}
\usepackage{threeparttable}
\usepackage[colorlinks=true,linkcolor=blue,citecolor=blue,allcolors=blue]{hyperref}
\usepackage[figuresright]{rotating}

\journal{Physica A: Statistical Mechanics and its Applications}

\begin{document}

\begin{frontmatter}



\title{Evolutionary Games on Networks: Phase Transition, Quasi-equilibrium, and Mathematical Principles}

\author[a,b]{Jiangjiang Cheng}
\ead{chengjiangjiang@amss.ac.cn}

\author[c]{Wenjun Mei}
\ead{mei@pku.edu.cn}

\author[d]{Wei Su}
\ead{suwei@amss.ac.cn}

\author[b]{Ge Chen\corref{cor1}}
\ead{chenge@amss.ac.cn}
\cortext[cor1]{Corresponding author.}

\address[a]{School of Mathematical Sciences, University of Chinese Academy of Sciences, Beijing, 100049,
China}

\address[b]{Key Laboratory of Systems and Control, Academy of Mathematics and Systems Science, Chinese Academy of Sciences, Beijing, 100190, China}

\address[c]{Department of Mechanics and Engineering Science, Peking University, Beijing, 100871, China}

\address[d]{School of Mathematics and Statistics, Beijing Jiaotong University, Beijing, 100044, China}

\begin{abstract}
    The stable cooperation ratio of spatial evolutionary games has been widely studied using simulations or approximate analysis methods. However, sometimes such ``stable'' cooperation ratios obtained via approximate methods might not be actually stable, but correspond to quasi-equilibriums instead. We find that various classic game models, like the evolutionary snowdrift game, evolutionary prisoner's dilemma, and spatial public goods game on square lattices and scale-free networks, exhibit the phase transition in convergence time to the equilibrium state. Moreover, mathematical principles are provided to explain the phase transition of convergence time and quasi-equilibrium of cooperation ratio. The findings explain why and when cooperation and defection have a long-term coexistence.
\end{abstract}



\begin{keyword}
    Networked evolutionary game \sep Phase transition \sep Quasi-equilibrium \sep Markov process
\end{keyword}

\end{frontmatter}


\section{Introduction}
\label{A}

Evolutionary game theory has become one of the major methodologies to study the stable equilibria of evolution of nature and human society \cite{0Martin, SZABO200797}. Since the network reciprocity has been considered as one of main mechanisms accounting for the evolution of cooperation \cite{nowak2006five}, the spatial evolutionary game describing the dynamics of spatially structured populations has received increasing attention in recent decades \cite{Ohtsuki2006A, 2005Evolutionary, ichinose2018mutation, braga2022stochasticity}. The research of spatial evolutionary game usually focuses on problems such as: What stable proportion are the cooperators going to converge to? This problem is quite complex because the stable proportion of cooperators depends on network topologies and game rules.

There are many previous works investigating the influence of network structures and game mechanisms on the stable cooperation ratio. Nowak and May studied the issue for evolutionary prisoner’s dilemma (EPD) on lattice networks using simulations and showed how cooperators resisted the invasion of defectors \cite{1992Evolutionary, 1993THE}. Santos and Pacheco confirmed via simulations that the scale-free network could enhance the stable cooperation ratio for both EPD and evolutionary snowdrift game (ESG) \cite{2005Scale}. Hauert and Doebeli demonstrated using simulations that spatial structure frequently inhibited the stable cooperation ratio in ESG compared to well-mixed populations \cite{2004Spatial}. Xu \textit{et al.} \cite{XU2022127698} verified through simulations that the cooperative behavior of multi-player ESG on scale-free simplicial complexes could be facilitated under some special parameter settings. Qin \textit{et al.} \cite{2008Effect} and Ren and Wang \cite{2014Robustness} showed via simulations that introducing memory effects promoted the stable cooperation ratio in EPD. Javarone proposed a model based on the kinetic theory of gases, and showed how motion could enhance the stable cooperation ratio for EPD \cite{EPJB2016}. Amaral and Javarone also studied the relationship between heterogeneity and stable cooperation ratio by implementing small and local perturbations on the payoff matrix of EPD \cite{PRSA2020}. Szab{\'o} and Hauert demonstrated an effective mechanism promoting the stable cooperation ratio by allowing voluntary participation in spatial public goods games (SPGG) \cite{szabo2002phase}. Li \textit{et al.} \cite{li2020evolution} found that time-varying network structure generally enhanced the evolution of cooperation in various social dilemmas. Perc \textit{et al.} \cite{PERC20171} mentioned the dynamic networks, in which each individual decided its neighbors at each time, could effectively improve cooperation. Xia \textit{et al.} \cite{Xia_2022} showed through simulations that costly reputation building in the trust game could promote the collective trust and cooperation within the networked population. Moreover, Assaf and Mobilia studied the influence of complex graphs on the stable cooperation ratio of ESG through mean-field dynamics \cite{2012Metastability}.

As mentioned above, various works of literature observe certain stable cooperation ratio using simulations or approximate theoretic methods. However, sometimes, the acquired ``stable'' cooperation ratio might not be actually stable. Namely, some interesting and fundamental questions have long been ignored: How long do the observed ``stable'' cooperation ratio last? Can numerical simulations mislead us on what the real equilibria are? This paper studies the convergence process of the ESG, EPD, and SPGG on square lattices and scale-free networks. We find phase transitions in convergence time to the equilibrium state. Futhermore, the study discovers that the average proportion of cooperators might be in quasi-equilibrium for a long time before reaching the real equilibrium state. Finally, from the perspective of absorbing Markov chains, mathematical principles are provided to explain the phase transition of convergence time and quasi-equilibrium of average cooperation ratio. Generally, the phase transitions are widely investigated in many stochastic multi-particle systems, such as contact process, directed percolation and Moran process \cite{marro_dickman_1999, AIP2000}. However, the phase transition of convergence time in spatial evolutionary games has never been studied, and there are a few literatures analyzing the convergence time based on the assumptions that the network topology is completed and the players' strategies are well-mixed \cite{2012Mixing, 2004Mathematical, 2006Fixation}. This study takes into account the topology structures of networks and spatial distributions of strategies. We find and analyze the phase transition of convergence time to the equilibrium state in spatial evolutionary games for the first time, presumably. These findings could give researchers some inspiration in finding and judging the real equilibria of spatial evolutionary games.

This paper is organized as follows. In Sec.~\ref{B}, we present two typical network structures and three typical spatial evolutionary game models. Sec.~\ref{C} presents some of the most relevant works and our research directions inspired by these works. In Sec.~\ref{D}, We find that there are two common phenomena in spatial evolutionary games: the phase transition of average convergence time and quasi-equilibrium of average cooperation ratio. In Sec.~\ref{E}, from the perspective of absorbing Markov chains, we give the mathematical principles behind these two phenomena. Finally, the conclusion and outlook are made in Sec.~\ref{F}.

\section{The models \label{B}}

This paper considers two typical networks: square lattice network and scale-free network. The size of the square lattice network is $\sqrt{N}\times\sqrt{N}$, with $N$ nodes and periodic boundary conditions \cite{2005Evolutionary}. The degree distribution of the scale-free network obeys the power law distribution $d(k)\sim k^{-\gamma}$, with the exponent $\gamma$ typically satisfying $2\leq\gamma\leq 3$. Its initial number of nodes, the number of edges per connection and the total number of nodes are $m_0$, $m(\leq m_0)$, $N$, respectively \cite{1999Albert}. Throughout this paper we choose $m=m_0=2$. For each node in both networks, the nodes directly connected to it are defined as its neighbors.

Three typical evolutionary games on networks are considered: the ESG, EPD and SPGG.

For the ESG and EPD, a population of individuals is arranged on a lattice or a scale-free network \cite{2005Scale}, in which each node represents a player with two strategies: cooperation ($C$) and defection ($D$), note that it repeatedly plays the snowdrift game or prisoner's dilemma with each of its neighbors. The symmetric payoff matrices for the snowdrift game and prisoner's dilemma are shown in the following tables 
\begin{figure}[htb]
    \centering
    \begin{tikzpicture}
         \node at (0, 1) {\bordermatrix {
                   & C & D \cr
                 C & 1-\frac{c}{2} & 1 - c \cr
                 D & 1 & 0
              }};
         \node at (0, 0) {Snowdrift game};
         \node at (4, 1) {\bordermatrix {
                   & C & D \cr
                 C & c & 0 \cr
                 D & 1 & 0
              }};
         \node at (4, 0) {Prisoner’s dilemma};
    \end{tikzpicture}
\end{figure}
where payoff matrices are rescaled, such as each one depends on the parameter $c\in(0,1)$ exclusively. We point out that when the elements in the payoff matrices of the two games are other numbers, the phase transition of convergence time and the quasi-equilibrium phenomenon of cooperation ratio mentioned later could still exist. In each round, every node $x$ calculates its total payoff $P_x$: the sum of the game's payoffs between node $x$ and every one of its neighbors. Apparently, $P_x$ depends on both node $x$'s strategy and the number of its neighbors choosing strategies $C$ and $D$ respectively.
\begin{figure}[t]
    \centering
    \includegraphics[width=\textwidth]{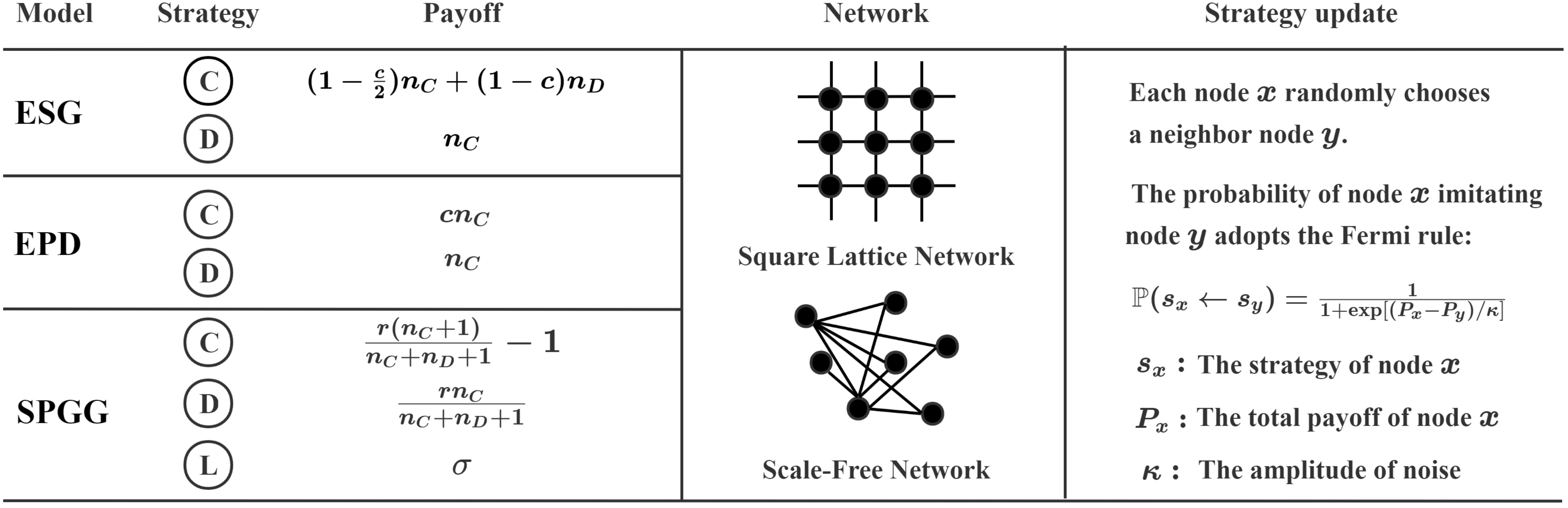}

    \centering
    \caption{\label{fig1} The diagram of spatial evolutionary games considered in this paper, where the ``Payoff'' column is the total payoff $P_x$ of any node $x$ when it chooses $C$, $D$, or $L$ in any time step, and $n_C$ (or $n_D$) represents the number of cooperative (or defected) neighbors. Note that the neighbors of each node do not include itself, so $n_C+n_D=4$ for the ESG and EPD on the square lattice network.}
\end{figure}

For the SPGG, the voluntary public goods game \cite{szabo2002phase} on the lattice or scale-free network is considered. Each node $x$ has three strategies: $C$, $D$, and $L$ (loneliness, i.e. not participating in the game), and in each round, its payoff $P_x$ is determined by a single public goods game played with all its neighbors. Specifically, each cooperator's contribution is normalized to unity, while defectors and loners contribute nothing. The sum of all the contributions in the group is multiplied by the enhancement factor $r>1$ as the total payoff and is shared equally among all the participants. The loner does not participate in the game and has a minor but reliable payoff $\sigma\in (0,r)$. Throughout this paper we choose $\sigma=1$.

Nodes in our models update their strategies via imitation-based rules. Assume each node in the above models randomly and uniformly chooses initial strategy from the set $\{C,D\}$ (or $\{C,D,L\}$). For the strategy update, each node $x$ randomly selects one of its neighbors $y$, and imitates node $y$'s strategy $s_y$ with a probability $\mathbb{P}(s_x\leftarrow s_y)$. Following previous works \cite{szabo1998evolutionary, 2018Mean,donahue2020evolving,2004Spatial,2004Emergence,szabo2002phase,hilbe2018evolution,amaral2018heterogeneous}, this paper adopts the widely used Fermi imitation rule:
\begin{equation*}
     \mathbb{P}(s_x\leftarrow s_y)=\frac{1}{1+\exp[(P_x-P_y)/\kappa]},
\end{equation*}
where $\kappa\geq 0$ represents the amplitude of the noise. When $\kappa\rightarrow 0$, node with higher payoff is always imitated; When $\kappa\rightarrow\infty$, the update process is dominated by randow drift.
Figure~\ref{fig1} provides a summary of the evolutionary games considered in this paper.

\section{Related works and motivation \label{C}}

The ESG, EPD and SPGG on different networks have been extensively studied in recent decades. For example, Nowak \textit{et al.} studied EPD on square lattice networks through simulations \cite{1992Evolutionary, 1993THE}, where each node imitated the strategy of the neighbor with the highest payoff, and the stable frequency of cooperators was obtained by time-averaging. Also, they presented an analytic approach for EPD on cycles and derived exact conditions for cooperative dominance \cite{ohtsuki2006evolutionary}.
Szab{\'o} and T{\H{o}}ke studied the cooperation ratio of EPD on a square lattice in the stationary state, and observed its continuous transition through Monte Carlo simulations and mean-field techniques \cite{szabo1998evolutionary}.
Santos and Pacheco studied ESG and EPD on regular ring graph and scale-free networks through simulations \cite{2005Scale}, which employed the proportional imitation rule to update the strategy of each node. Stable proportion of cooperators was obtained by averaging over 1000 generations after a transient time of 10,000 generations. 
Chiong and Kirley studied the effects of random mobility on the cooperative evolution of EPD on a square lattice, and they showed that random mobility improves the level of cooperation compared to the static model \cite{chiong2012random}.
Santos \textit{et al.} studied the EPD on square lattice networks by introducing noise in the decision making process, and observed three stable states: cooperators and defectors absorbing states and a coexistence state between them \cite{santos2017phase}.
\begin{figure}[htb]
    \centering
    \includegraphics[width=\textwidth]{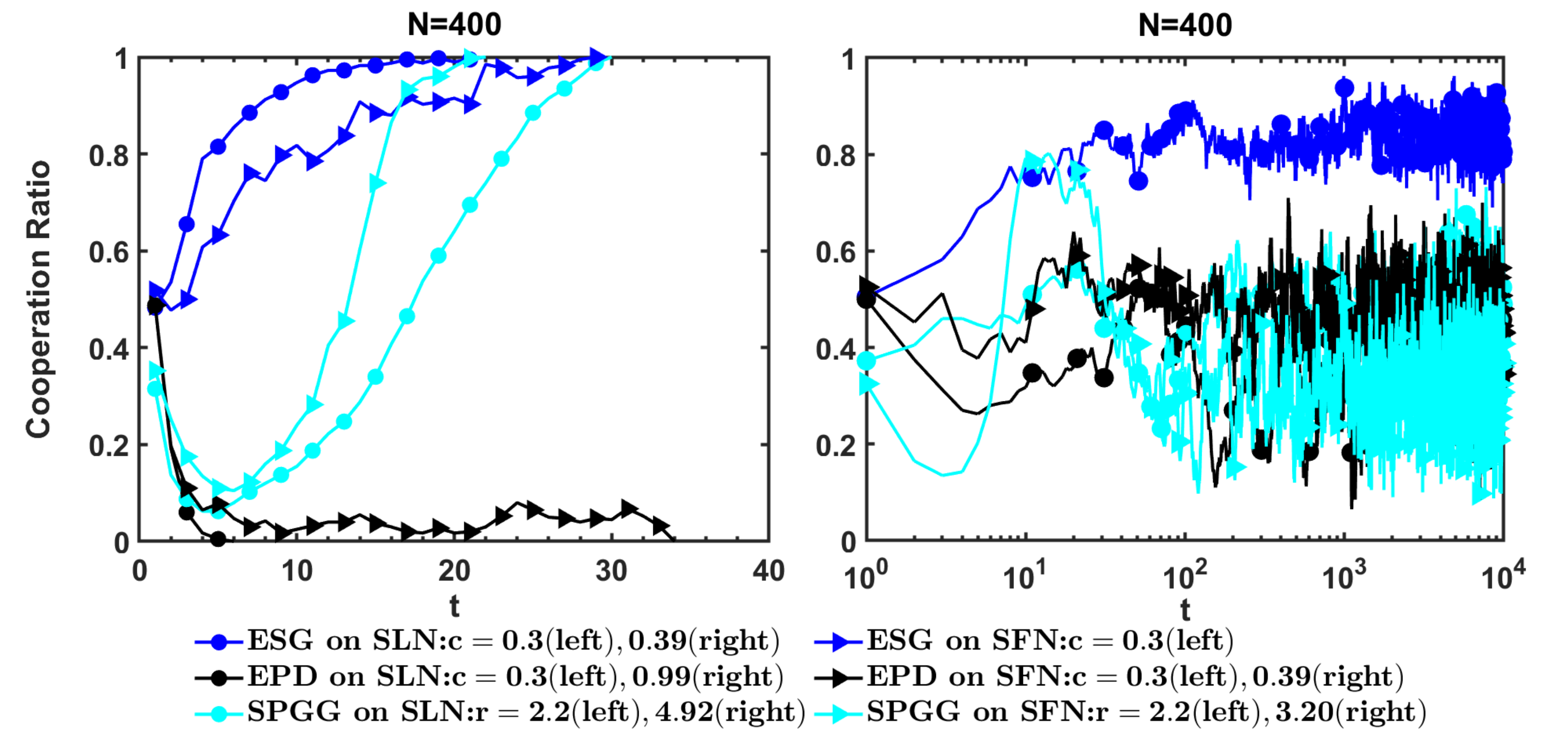}
    \caption{ The curves of the cooperation ratio of spatial evolutionary games on SLN (square lattice network) and SFN (scale-free network)($\kappa=0.1$).}
    \label{fig2}
\end{figure}
Boccaletti \textit{et al.} summarized the findings of the emergence of cooperation in the EPD, ESG and SPGG in multilayer networks, where each individual played with its neighbors at the same network layer and different network layers \cite{boccaletti2014structure}. Flores \textit{et al.} \cite{FLORES2022112744} explored the cooperative survival of EPD and SPGG on several regular lattices by using both analytical methods and agent-based Monte Carlo simulations.
Hauert and Doebeli considered ESG on regular lattice networks using proportional imitation rule and Fermi rule, respectively \cite{2004Spatial}. The stable proportion of cooperators were determined by evolving the lattice over 10,000 generations and then averaging over another 1000 generations. 
Qin \textit{et al.} \cite{2008Effect} and Ren and Wang \cite{2014Robustness} studied EPD on square lattice networks by introducing memory effects. The stable proportion of cooperators was similarly approximated through simulation averaging. Besides, the SPGG on square lattice networks was studied in \cite{szabo2002phase, 2014Probabilistic, 2015Competition, JOSM2016}, using the Fermi imitation rule and obtaining the stable cooperation ratio through Monte Carlo simulations. However, almost all studies ignored a fundamental problem: can the ``stable cooperation ratio'' observed by simulations last forever? If not, what is the long-term trend of the cooperation ratio? Can we find the mathematical principle for the long-term behavior of cooperation ratio? Exploring these issues constitutes the first motivation of this paper.

On the other hand, the convergence time (fixation time) is also a very important issue. Black \textit{et al.} \cite{2012Mixing} approximated the mixing times of evolutionary games under the assumptions of complete graphs and well-mixed strategies, where the mixing time is the time of the probability
distribution over states to approach its stationary distribution. Assaf and Mobilia \cite{2012Metastability} approximated the mean fixation time of the ESG on scale-free networks by mean-field dynamics. Hajihashemi and Samani \cite{hajihashemi2019fixation} used an analytical method based on Markov chains to calculate the mean fixation time of the birth-death process on many evolutionary graph structures.
Differently from these works, we take into account the topology structures of networks and spatial distributions of strategies, and then explore the average convergence time of spatial evolutionary games with respect to system parameters by both theoretical analysis and numerical experiments.

\section{Phase transition and Quasi-equilibrium \label{D}}

\subsection{Phase transition of convergence time}

In general, the above models eventually converge to either total cooperation or total defection. In many situations the convergence to total cooperation or total defection is fast (see the left panel of Figure~\ref{fig2}), however in other situations the systems may exhibit long-term coexistence of cooperation and defection, see the right panel of Figure~\ref{fig2}, and the average of 1,000 simulations is shown in Figure~\ref{fig4}.
\begin{figure}[htb]
    \centering
    \includegraphics[width=\textwidth]{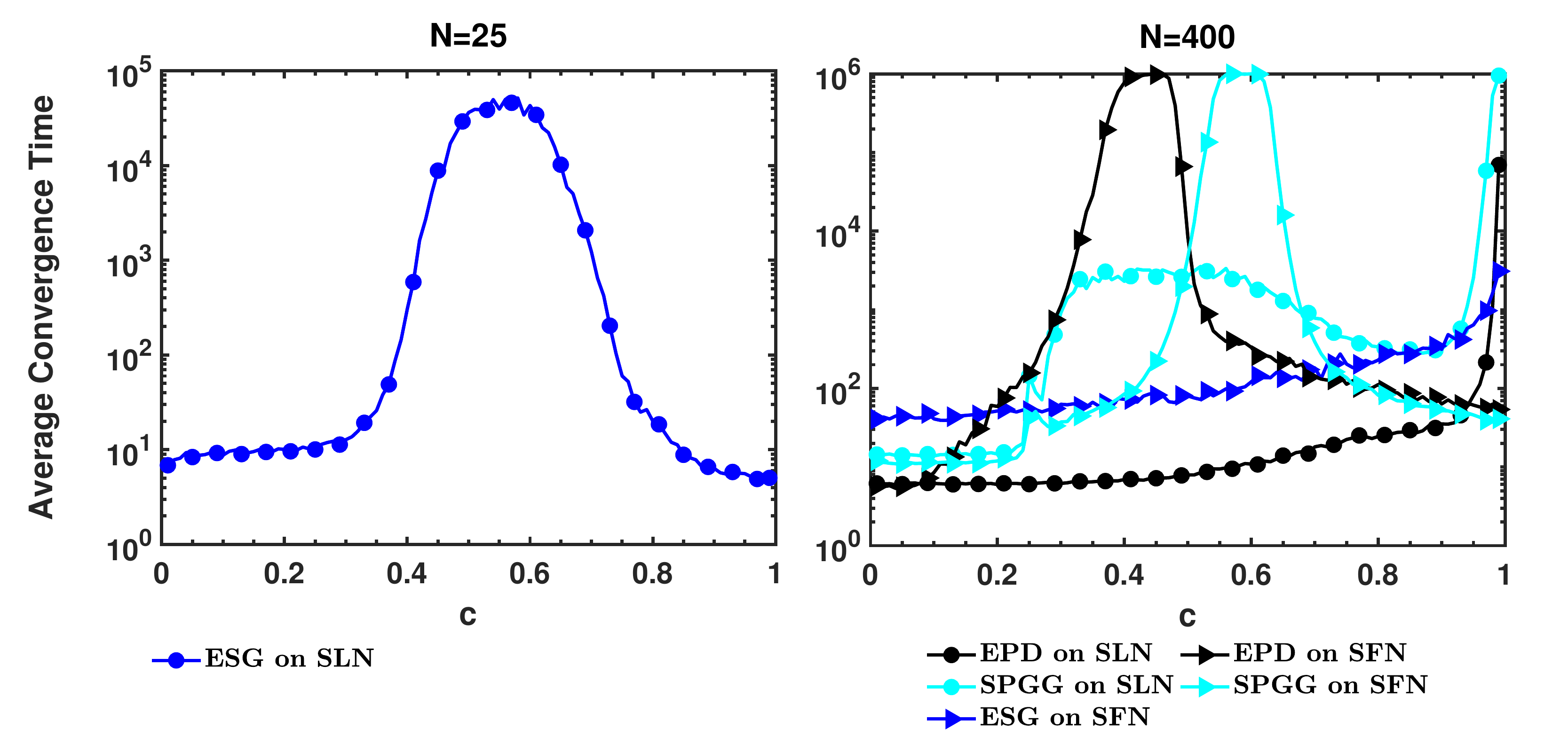}

    \centering
    \caption{\label{fig3} The curves of average convergence times to equilibrium state of spatial evolutionary games with 100 repetitions ($\kappa=0.1$). For the SPGG, in order to unify the parameters, we set $c=\frac{r-1}{4}(0<c<1)$.}
\end{figure}
\begin{figure}[H]
    \centering
    \includegraphics[width=0.68\textwidth]{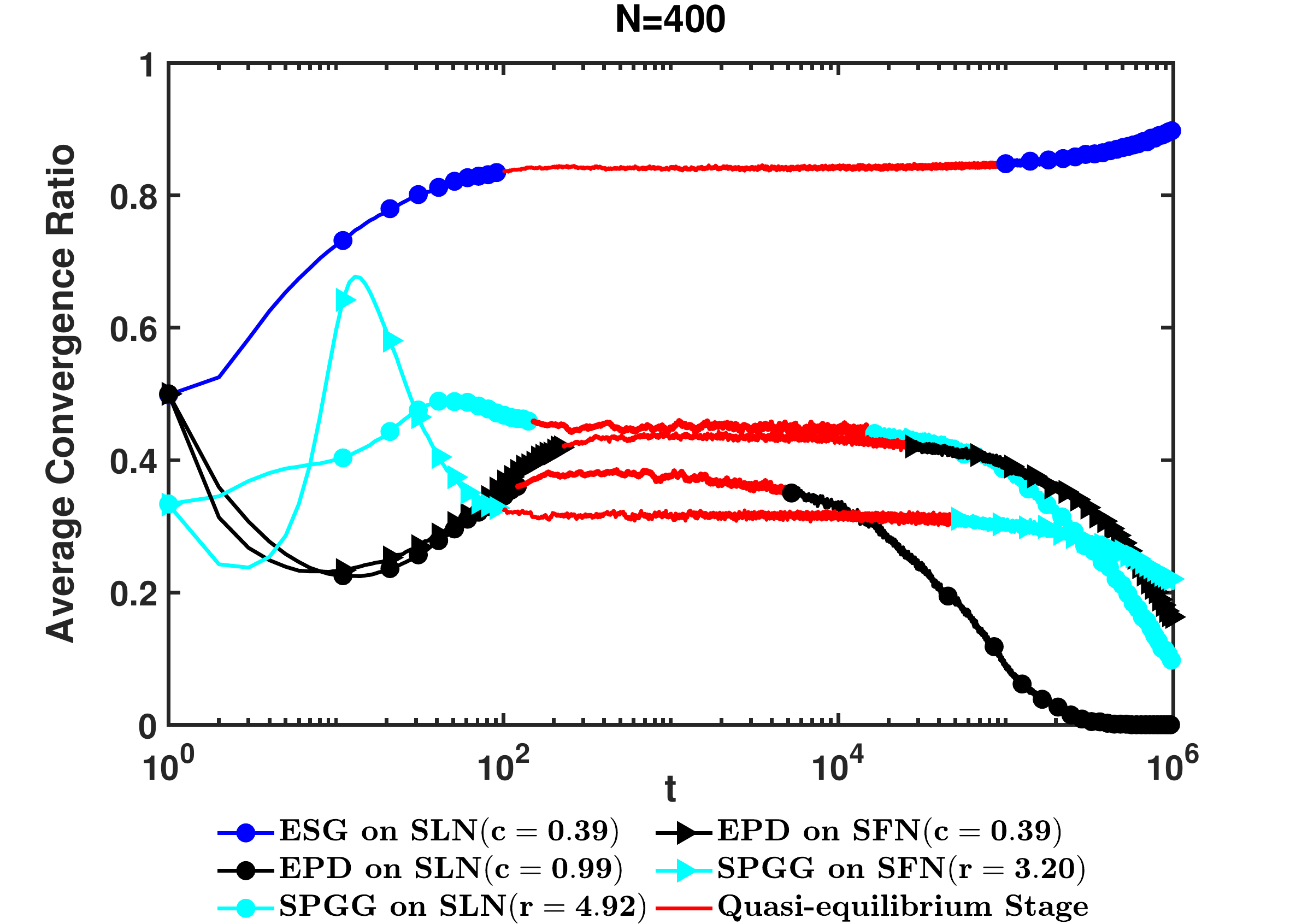}

    \centering
    \caption{\label{fig4} The curves of average cooperation ratios of spatial evolutionary games with 1,000 repetitions ($\kappa=0.1$). The red part of each curve denotes the quasi-stable stage.}
\end{figure}

The speed of the convergence to equilibria plays an important role for the coexistence time of cooperation and defection. However, the importance of convergence speed has been ignored, usually. In this setting, this study comprehensively investigates the speeds of the convergence to equilibria of spatial evolutionary games and finds the phase transition for the ESG, EPD, and SPGG on the square lattice network, and EPD and SPGG on the scale-free network, as shown in Figure~\ref{fig3}. Because the convergence time changes sharply concerning the parameter $c$ at some critical points, we speculate that there exist phase transition in the convergence time if the network size grows to infinite. Note that the ordinate in Figure~\ref{fig3} adopts a logarithmic scale. The phase transition is clearly more observed when using the ordinary coordinate system, see Figure~\ref{fig7} as an example. The phase transition of ESG on the square lattice network is the most obvious, so we only show the case when $N=25$. It can be seen from Figure~\ref{fig3} and Figure~\ref{fig7} that when the population size increases, the time of the convergence to equilibria is astronomical under some parameter conditions, which means the cooperation and defection coexist for an extremely long period. This finding explains why and when the cooperation and defection have a long-term coexistence. It is worth mentioning that the anomalous increase of extinction time has also been investigated in different systems, such as the domain growth yields slow relaxation and the Griffith phase occurring for quenched randomness in spatial evolutionary games \cite{SZABO200797}.

\subsection{Quasi-equilibrium}

In this context, an interesting problem is posed: What is the trend of the long-term coexistence of cooperation and defection? From the right panel of Figure~\ref{fig2},  the cooperation ratio exhibits a large fluctuation. However, when calculating the average value of the cooperation ratio with 1,000 repetitions, this study discovers that it has a long-term quasi-stable stage prior to reaching the equilibrium states. This is called a quasi-stable stage in similarity with quasi-equilibrium, see Figure~\ref{fig4} (note that the horizontal axis in the figure adopts a logarithmic scale).
Figure~\ref{fig4} shows that the average cooperation ratio might stay in quasi-equilibrium for an extended period and gradually reaches the equilibrium state afterwards. Moreover, when the population size becomes large, the duration of the quasi-equilibrium might become astronomical. It is worth mentioning that due to the huge duration of quasi-equilibrium, there are many previous works treating the quasi-equilibrium as the real stable cooperation ratio \cite{2005Scale, 2004Spatial, 2008Effect, 2014Robustness, szabo2002phase, szabo1998evolutionary, santos2017phase, 2014Probabilistic, 2015Competition, JOSM2016}.

Another interesting problem is the question of what is the mathematical principle behind the above-mentioned phase transition and quasi-equilibrium phenomena. The following will take the ESG on the square lattice as an example to study this issue. Note that other types of evolutionary games considered in this study are essentially the same.

\section{Mathematical principles \label{E}}

We first introduce the basic content of Markov chain \cite{1960Markov}. Let $(\Omega,\mathcal{F},\mathbb{P})$ be a probability space, $\mathcal{S}$ be a finite state space, and $\mathcal{T}=\{0,1,2,\cdots\}$ be the time domain. Assume $\{X_t,t\in\mathcal{T}\}$ is a sequence of states, which is also a sequence of
random variables defined in $(\Omega,\mathcal{F},\mathbb{P})$. If $\{X_t,t\in\mathcal{T}\}$ satisfies
\begin{equation*}
     \mathbb{P}(X_{n}=i_n|X_{0}=i_0,\cdots,X_{n-1}=i_{n-1})\\
     =\mathbb{P}(X_{n}=i_n|X_{n-1}=i_{n-1}), \forall n\geqslant 1, i_0,\cdots,i_n\in\mathcal{S},    
\end{equation*}
we say $\{X_t,t\in\mathcal{T}\}$ is a discrete time Markov chain with finite state. Let
\begin{equation*}
     M_{ij}(t):=\mathbb{P}(X_{t+1}=j|X_{t}=i),\forall i,j\in\mathcal{S}, \forall t\in\mathcal{T},
\end{equation*}
and then $M(t)=[M_{ij}(t)]$ is the transition probability matrix of the Markov chain $\{X_t,t\in\mathcal{T}\}$ at time $t$. If 
\begin{equation*}
     M(0)=M(1)=M(2)=\cdots=M,
\end{equation*}
we say $\{X_t,t\in\mathcal{T}\}$ is a time-homogeneity Markov chain with the transition probability matrix $M$.

According to the ESG's strategy update rule, it is clear that the ESG belongs to a time-homogeneity Markov chain with finite state, and its state space on $\sqrt{N}\times\sqrt{N}$ square lattice can be defined as $\{C,D\}^{\sqrt{N}\times\sqrt{N}}$, then the total number of states is $2^{N}$. For the convenience of discussion, we number all elements of $\{C,D\}^{\sqrt{N}\times\sqrt{N}}$ by $\mathcal{E}:=\{1,2,\ldots,2^{N}\}$. Also, the ESG has two equilibrium states:  total cooperation and total defection, which constitute the absorption states of Markov chain.
Let  $\mathcal{A}$ and $\mathcal{N}$ denote the number sets of absorption states and transition (non absorption) states of the ESG respectively, then $\mathcal{E}=\mathcal{A}\cup\mathcal{N}$.
\begin{figure}[H]
    \centering
    \includegraphics[width=0.68\textwidth]{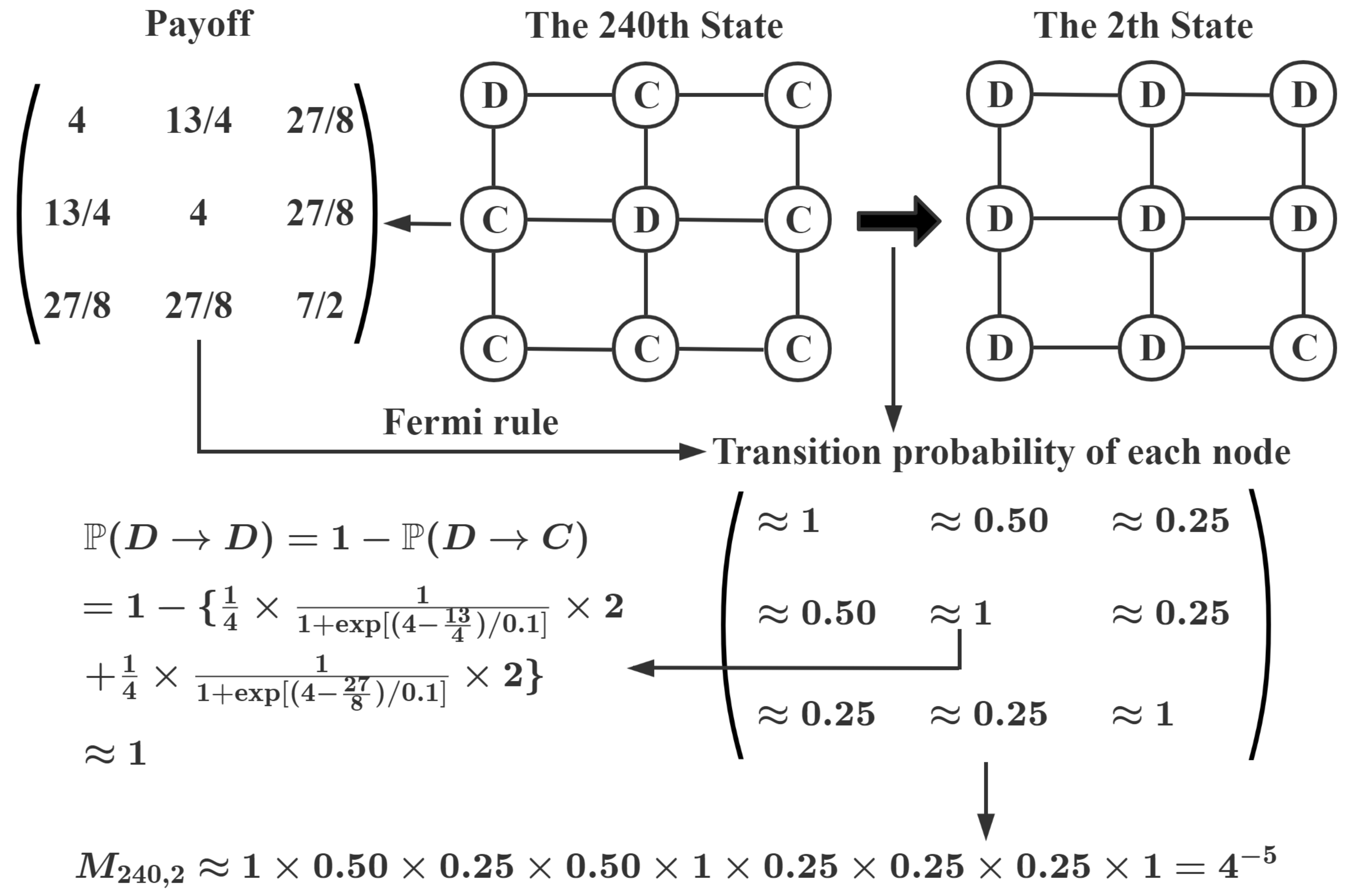}

    \centering
    \caption{\label{fig6} The calculation of the element $M_{240,2}$ in the transition probability matrix $M(1/4,9,0.1)$ for the ESG with $c=1/4, N=9, \kappa=0.1$.}
\end{figure}

We define the transition probability matrix $M=M(c,N,\kappa)\in[0,1]^{2^{N}\times 2^{N}}$, whose element $M_{ij}$ denotes the transition probability from state numbers $i$ to $j$.
Figure~\ref{fig6} shows an example to calculate the elements of the transition probability matrix $M$. The following table shows the effect on the element $M_{240,2}$ of the transformation probability matrix when $\kappa$ adopts different values.
\begin{table}[H]
	\centering
	\begin{tabular}{l|cccccccccc}  
		\hline
		$\kappa$ &0.01 &0.05 &0.1 &0.3 &0.6 &1 &3 &5 &7 &10\\
		\hline
        $M_{240,2} (.\times 10^{-4})$ &$9.77$ &$9.77$ &$9.66$ &$4.29$ &$1.01$ &$0.36$ &$0.09$ &$0.06$ &$0.06$ &$0.05$\\
		\hline
	\end{tabular}
    \caption*{The element $M_{240,2}$ in the transition probability matrix for the ESG with $c = 1/4, N = 9$ when $\kappa$ adopts different values.}
\end{table}
Let $T(M)$ be the expectation of reaching time from an initial state randomly and uniformly chosen in $\mathcal{E}$ to absorption states $\mathcal{A}$.
According to the results of absorbing Markov chains \cite{kijima2013markov}, we have $T(M)=\frac{1}{2^{N}}\mathbf{1}^{\top}(I-M^\ast)^{-1}\mathbf{1}$, where $I$ is the $(2^{N}-2)$-order identity matrix, $M^\ast\in[0,1]^{(2^{N}-2)\times (2^{N}-2)}$ is the remaining matrix of that $M$ deletes the corresponding rows and columns of its absorption states, and $\mathbf{1}=(1,\cdots,1)^{\top}$.

Using the power series expansion of matrix functions, we have
\begin{equation*}
    T(M)=\frac{1}{2^{N}}\mathbf{1}^{\top}(I-M^\ast)^{-1}\mathbf{1}
    =\frac{1}{2^{N}}\sum_{k=0}^{\infty}\mathbf{1}^{\top} {M^{\ast}}^k\mathbf{1}
    =\frac{1}{2^{N}}\sum_{k=0}^{\infty}\sum_{i=1}^{2^{N}-2}\sum_{j=1}^{2^{N}-2}({M^\ast}^k)_{ij},
\end{equation*}
then according to the theory of Markov chains \cite{kijima2013markov}, we have the following estimation:
\begin{equation}
     T(M)\leq\frac{1}{2^{N}}\sum_{k=0}^{\infty}\sum_{i=1}^{2^{N}-2}\sum_{j=1}^{2^{N}-2}C_N\rho^k(M^{\ast})
     =\frac{C_N (2^{N}-2)^2}{2^{N}}\sum_{k=0}^{\infty}\rho^k(M^{\ast})
     =O\left(\frac{1}{1-\rho(M^{\ast})}\right),
     \label{2}
\end{equation}
where $C_N=O(\frac{1}{2^{N}-2})$ \footnotemark[1] is a constant, and $\rho(M^{\ast})\in(0,1)$ is the spectral radius of matrix $M^{\ast}$.

\footnotetext[1]{$C_N=O(\frac{1}{2^{N}-2})$ means that there exists a constant $b>0$ and an integer $N_0>0$ satisfying $C_N\leq \frac{b}{2^{N}-2}$ for all $N\geq N_0$.}

\begin{figure}[htb]
    \centering
    \includegraphics[width=0.83\textwidth]{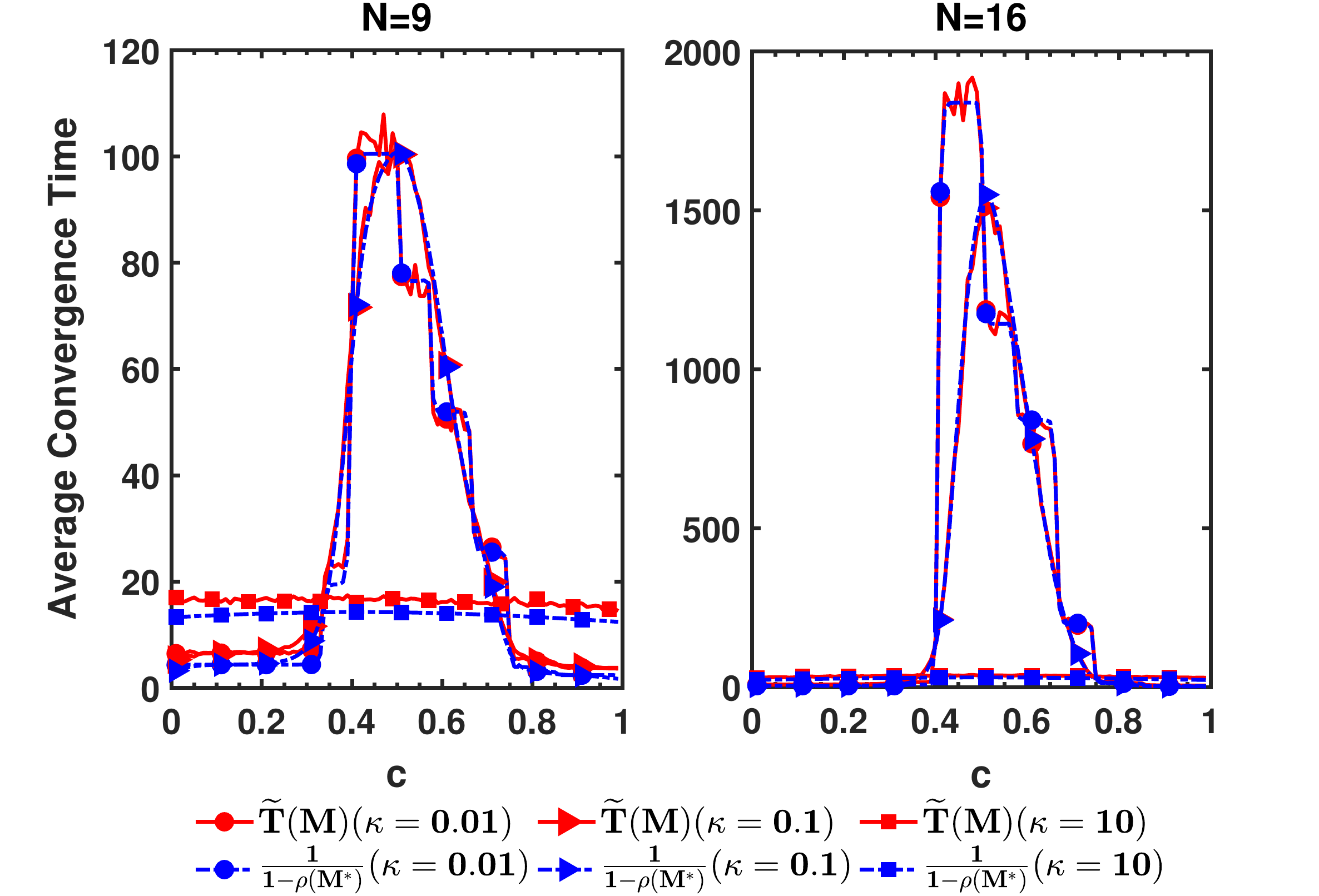}

    \centering
    \caption{\label{fig7} The curves of $\frac{1}{1-\rho(M^{\ast})}$ (blue) and the average convergence time $\widetilde{T}(M)$ with 1,000 repetitions (red) under the ESG when $N=9$ and $N=16$.}
\end{figure}
Eq.~(\ref{2}) gives that $T(M)$'s upper bound has the same order of  $\frac{1}{1-\rho(M^{\ast})}$. In fact, Ref.~\cite{saglam2014metastable} shows that when initial states satisfy a certain probability distribution, $T(M)=\frac{1}{1-\rho(M^{\ast})}$. We make a comparison between $T(M)$ and $\frac{1}{1-\rho(M^{\ast})}$ through simulations. In detail, we carry out the ESG on the square lattice network with $N=9, 16$ and $\kappa=0.01, 0.1, 10$, and use the average convergence time $\widetilde{T}(M)$ with 1,000 repetitions to approximate $T(M)$.  On the other hand, we compute the transition probability matrix $M$ and the spectral radius $\rho(M^{\ast})$. To make a comparison, Figure~\ref{fig7} draws the curves of $\widetilde{T}(M)$ and $\frac{1}{1-\rho(M^{\ast})}$ with respect to $c$. From Figure~\ref{fig7}, it can be clearly obtained that the average convergence time $\widetilde{T}(M)$ is close to $\frac{1}{1-\rho(M^{\ast})}$, and the phase transition happens for both $\widetilde{T}(M)$ and $\frac{1}{1-\rho(M^{\ast})}$ when the noise amplitude $\kappa\rightarrow 0$.  It is conjectured that the phase transition of convergence time is caused by change of $\rho(M^{\ast})$.  When $\rho(M^{\ast})$ tends to $1$, the corresponding convergence time increases sharply, and the phase transition occurs.
\begin{figure}[htb]
    \centering
    \includegraphics[width=0.68\textwidth]{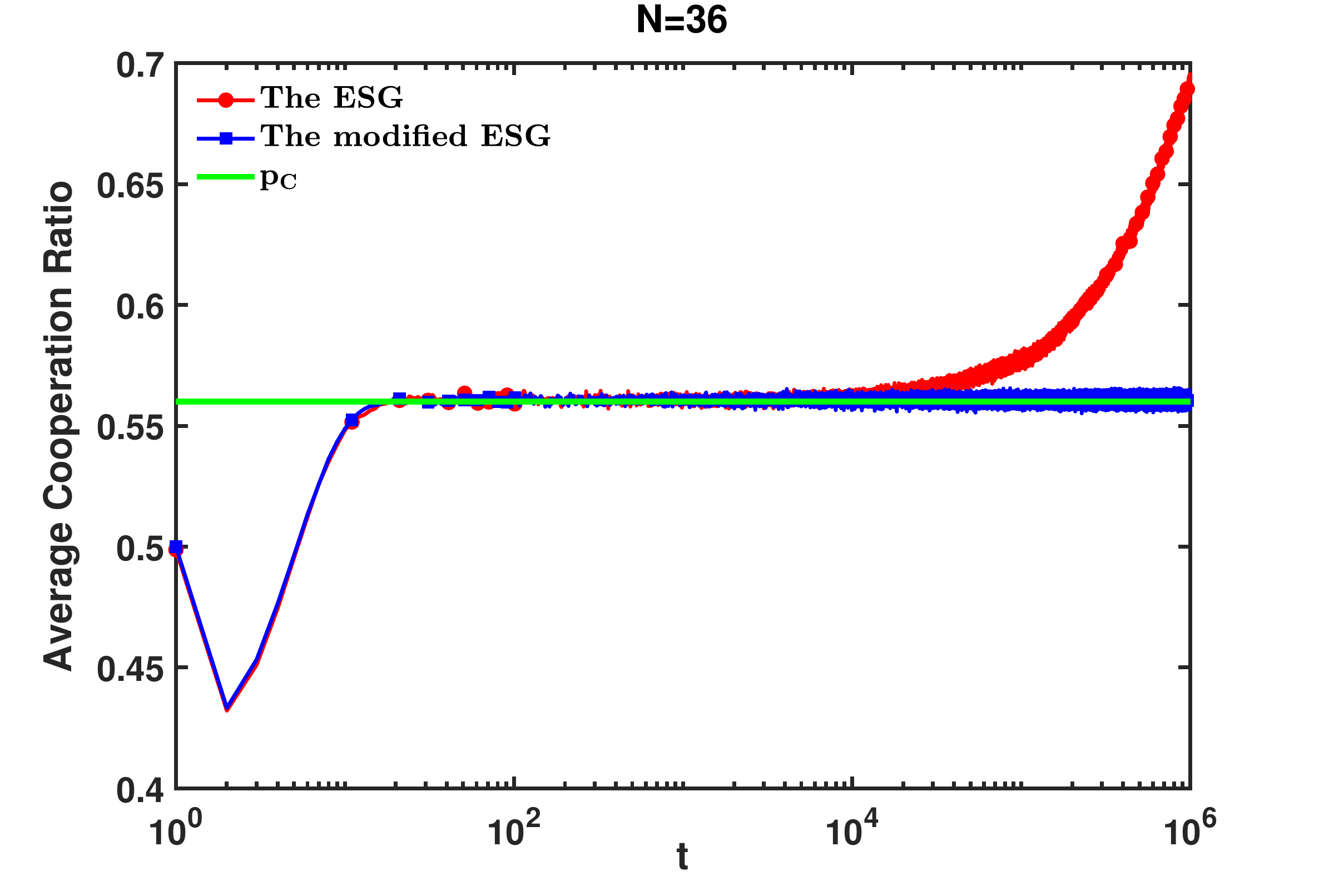}

    \centering
    \caption{\label{fig8} The line of $p_C$ (green), and the curves of the average cooperation ratios with 1,000 repetitions under the ESG (red) and modified ESG (blue) when $c=0.58, N=36, \kappa=0.1$.}
\end{figure}

On the other hand, the quasi-equilibrium can be derived by stationary distribution of the  transition probability matrix $\widetilde{M}\in[0,1]^{(2^{N}-2)\times (2^{N}-2)}$, which is
formed by adding diagonal elements of $M^{\ast}$ such that the sum of its each row equals $1$. Note that $\widetilde{M}$ does not contain any absorption state. Let $\widetilde{x}$ denote the stationary distribution for $\widetilde{M}$, then clearly $\widetilde{x}=\widetilde{M}^\top\widetilde{x}$. Let $p_C:=\sum_{i\in\mathcal{N}}\widetilde{x}_iC(i)$, where $\widetilde{x}_i$ is the component of $\widetilde{x}$ with state number $i$, and $C(i)$ represents the proportion of cooperation in the state numbered $i$, that is, the number of cooperative nodes divided by the total number of nodes $N$. According to Ref.~\cite{saglam2014metastable}, the average cooperation ratio of the ESG in the quasi-stable stage can be approximated by $p_C$.

Since the order of $\widetilde{M}$ increases exponentially with the growth of $N$, the stationary distribution $\widetilde{x}$ is hard to calculate directly. As an alternative, we construct an evolutionary game according to $\widetilde{M}$, and then use the Monte Carlo simulation to approximate $p_C$. Specifically, the evolutionary game we construct is the same as the ESG, except that when the system state reaches total cooperation or defection, the system goes back to the previous state immediately. It can be verified that the transition probability matrix corresponding to the modified ESG  is exactly $\widetilde{M}$, and $p_C$ can be approximated by the limitation of the average cooperation ratio under the modified ESG.

Figure~\ref{fig8} shows the curves of the average cooperation ratios under the ESG and modified ESG when $c=0.58, N=36, \kappa=0.1$ and the corresponding theoretical cooperation ratio $p_C$ in quasi-stable stage (Note that the horizontal axis in the figure adopts a logarithmic scale). It can be seen from Figure~\ref{fig8} that the average cooperation ratio of the modified ESG converges to $p_C$ quickly, while the average cooperation ratio of the ESG reaches $p_C$ quickly and stays on $p_C$ for a long time.

\section{Summary \label{F}}

This study considers three types of evolutionary games on both square lattice and scale-free networks and finds the phase transition phenomenon in convergence time to equilibrium states and the quasi-equilibrium of average cooperation ratio. Furthermore, from the perspective of absorbing Markov chains, multiple mathematical principles that are behind them are given. Various interesting works are still waiting for further discussion. For example, the formulas of convergence time  and  quasi-equilibrium with respect to the network's size and game parameters might be studied, providing better theoretical guidance on how to estimate and enhance the cooperation ratio of spatial evolutionary games.

\section{Acknowledgments}

This research is supported by the National Key Research and Development Program of China (2022YFA1004600),
the Strategic Priority Research Program of Chinese Academy of Sciences (XDA27000000),
the National Natural Science Foundation of China (72192800, 12288201, 72201008, 12071465, 72131001),
and the Fundamental Research Funds for the Central Universities (2021RC267).

The computations were (partly) done on the high performance computers of State Key Laboratory of Scientific and Engineering Computing, Chinese Academy of Sciences.



\bibliographystyle{elsarticle-num} 
\bibliography{main}





\end{document}